\documentclass[a4paper, 12pt]{article}
\usepackage{amsmath,amsfonts,amsthm, mathtools}
\usepackage{txfonts}
\usepackage{amssymb}
\usepackage{fullpage,url, hyperref}
\usepackage{longtable, threeparttable, booktabs, graphicx, float}
\usepackage{times, txfonts, color}
\usepackage{xcolor}
\usepackage{titlesec}
\usepackage{algorithmic}
\usepackage[linesnumbered,ruled]{algorithm2e}
\usepackage{mathtools}
\usepackage{mathrsfs}
\usepackage{natbib}
\usepackage{caption}
\usepackage{subcaption}

\theoremstyle{definition}

\begin{document}

 \title{\bf A Bayesian bivariate conditional Poisson regression for goal dependence in the English Premier League}
 \author{Marcus Nolan$^\star$\footnote{\texttt{email}: \href{mailto:marcus.nolan1234@gmail.com}{marcus.nolan1234@gmail.com}}\,\,\,, Wagner Barreto-Souza$^\star$\footnote{\texttt{email}: \href{mailto:wagner.barreto-souza@ucd.ie}{wagner.barreto-souza@ucd.ie}},\,\, Luiza S.C. Piancastelli$^\star$\footnote{\texttt{email}: \href{mailto:luiza.piancastelli@ucd.ie}{luiza.piancastelli@ucd.ie}}\\ and  Raanju R. Sundararajan$^\sharp$\footnote{\texttt{e-mail}: \href{mailto:rsundararajan@mail.smu.edu}{rsundararajan@mail.smu.edu}}
\\
\normalsize \it $^\star$School of Mathematics and Statistics, University College Dublin, Republic of Ireland\\
\normalsize \it $^\sharp$Department of Statistics and Data Science, Southern Methodist University, USA}
 \maketitle

\begin{abstract}
	Understanding the relationship between home and away goal counts in football provides valuable insights into match-level dynamics. While the influence of home advantage is well-established, with historical records indicating roughly 50\% of matches are won by home teams (versus about 30\% by the away team), properly determining the joint goal distribution while accounting for key match factors remains under-explored. In this paper, we develop a Bayesian bivariate Conditional Poisson (BCP) regression model to explicitly capture the dependence between home and away goal counts, addressing a core limitation of traditional football scoring models that assume independence or only allow for positive correlation. The BCP regression is applied to the English Premier League (EPL) data spanning three seasons, incorporating stadium attendance and committed fouls by both teams as regressors. Inference is conducted within a Bayesian framework, enabling interpretable uncertainty quantification and model validation through posterior predictive checks. Our results reveal a negative correlation between home and away goal counts and highlight an asymmetry in how match attendance influences home versus away scoring.
\end{abstract}


\section{Introduction}
\label{sec:introduction}

Football scores are naturally represented by a pair of non-negative integer-valued outcomes: the goals scored by the home team and those scored by the away team. These counts determine the match result, but they also reflect a sequence of strategic interactions. A goal may alter the pace of play, the amount of risk taken by each team, and the balance between attacking and defensive effort. Consequently, home and away goals are not expected to be independent. Quantifying their dependence while accounting for match-level factors is important both for understanding match dynamics and for producing realistic probabilistic descriptions of football scores.

Poisson models provide a standard starting point for this problem. The seminal formulation of \citet{maher1982} expressed the scoring intensities of the two teams in terms of their attacking and defensive strengths and a home-effect parameter. Subsequent work relaxed the assumption that the two goal counts are independent. In particular, \citet{dixon1997} introduced an adjustment for dependence in low-scoring matches, while \citet{karlis2003} showed that bivariate Poisson models can improve the representation of draws and other score-line probabilities. These contributions established that the association between home and away goals is an intrinsic feature of football data. However, the conventional bivariate Poisson construction generates dependence through a shared Poisson component and therefore permits only non-negative correlation. This restriction may be inappropriate when in-match tactical responses produce negative dependence, for example when a team protects a lead by ``parking the bus", and suppresses its opponent's scoring opportunities.

The statistical relationship between home and away goals is also connected to the well-documented phenomenon of home advantage. Using English football data, \citet{clarke1995} demonstrated that home advantage should be assessed after accounting for differences in team strength, rather than solely through the percentage of home wins. International evidence further indicates substantial geographical variation in the magnitude of home advantage, consistent with a combination of crowd support, travel, familiarity, and territorial factors \citep{pollard2006}. Crowd support is of particular interest because it may affect both player performance and match officials. Experimental evidence reported by \citet{nevill2002} showed that crowd noise can influence referees' assessments of challenges, providing a plausible pathway through which spectators may affect fouls, disciplinary decisions, and ultimately scoring opportunities.

The attendance restrictions imposed during the COVID-19 pandemic created an unusual setting in which the role of spectators could be examined more directly. Across 17 countries and 23 leagues, \citet{bryson2021} found that matches played without crowds were associated with fewer disciplinary sanctions against away teams, thereby reducing one channel of home advantage. Similarly, \citet{scoppa2021} reported that the home advantage in goals, points, and other performance measures was substantially smaller in matches played behind closed doors, while foul and card decisions became more balanced between home and away teams. Evidence from Germany suggests that the effect was not uniform across competitions: \citet{fischer2021} found a reduction in home advantage in the top division but not in the second and third divisions. A broad analysis of European leagues by \citet{mccarrick2021} likewise documented an overall decline in home advantage without spectators, although the magnitude varied across leagues and outcomes.

More recent statistical developments have addressed the joint distribution of home and away goals. \cite{koolit2015} introduced a dynamic bivariate Poisson model to analyze English Premier League (EPL) match outcomes from the 2010–2011 and 2011–2012 seasons. \cite{piancastelli2023} developed a flexible multivariate COM-Poisson model for analyzing EPL matches and found that the absence of crowds was associated primarily with a reduction in home-team scoring, rather than a comparable change in away-team goals. This asymmetry is especially relevant for the present study: if attendance affects the two scoring processes differently, a model should permit separate covariate effects for home and away goals while simultaneously representing their dependence. Foul measures may provide additional information because they capture aspects of defensive pressure, interruptions of attacking play, and possible crowd-related differences in officiating. Other recent contributions on the analysis of the number of goals scored by home and away teams based on the COM-Poisson distribution are due to \cite{floetal2025} and \cite{zhaetal2026}. To extend the baseline bivariate Poisson literature, \cite{micetal2025} framed the Dixon-Coles model as a special case of the Sarmanov family and proposed flexible multiplicative extensions tailored to the distinct scoring dynamics of women's football. Extending multivariate count modeling beyond goal counts to match discipline, \cite{phi2026} proposed a bivariate mean-parameterized COM–Poisson copula model to evaluate referee heterogeneity and yellow card distributions across major European leagues.

To overcome the limitations of standard football scoring models, which often assume independence or restrict dependence to positive correlation, this paper proposes a flexible Bivariate Conditional Poisson (BCP) regression framework to model match-level dynamics and goal dependence. Following \cite{berkhout2004} and the reparameterization by \cite{piancastelli_ingarch}, the BCP specification accommodates an extended range of joint dependence, allowing for a comprehensive analysis of English Premier League match dynamics, stadium attendance, and foul-committed measures across pre-, during-, and post-pandemic seasons. A full Bayesian inferential framework is explicitly developed to estimate the model parameters via Hamiltonian Monte Carlo sampling. This procedure yields complete posterior distributions for both structural dependence and covariate parameters while facilitating rigorous model validation through posterior predictive checks. The main contributions are listed below. 
\begin{itemize}
    \item The flexible BCP formulation specifies one goal count marginally and the other conditionally, introducing a dependence parameter $\phi$ that captures both positive and negative associations.
    
    \item Stadium attendance and foul-related variables enter the marginal mean specifications as covariates, enabling separate estimation of their associations with home and away scoring while simultaneously modeling goal dependence.
    
    \item Inference is conducted within an empirical Bayes framework, providing posterior distributions for parameter uncertainty quantification and a natural basis for model evaluation via posterior predictive checks.
    
    \item The empirical analysis uses 1,140 matches from the 2018--19, 2020--21, and 2023--24 seasons, representing pre-pandemic, restricted-attendance, and post-pandemic conditions. The aims are (i) to determine whether goal counts retain structural dependence after accounting for attendance and foul variables, (ii) to assess whether attendance is associated differently with the two teams' scoring rates, and (iii) to evaluate whether the flexible dependence structure of the BCP model provides an informative description of EPL match outcomes.
\end{itemize}

In a closely related paper, \cite{petetal2025} proposed a bivariate conditional Poisson specification to model goal dependence in football matches. Although also inspired by \cite{berkhout2004}, their formulation adopts a more complex conditional structure for which key statistical properties, such as marginal expectation and variance of one component, and the resulting correlation, cannot be obtained in closed form. In contrast, our approach relies on a tractable, well-established parameterization where these moments and dependence metrics are analytically well-defined. A detailed discussion of these methodological distinctions is provided in Section~\ref{sec:methods}.

The remainder of this article is organized as follows. Section~\ref{sec:data} details the dataset of 1,140 English Premier League matches across pre-pandemic, restricted-attendance, and post-pandemic seasons, presenting descriptive statistics and preliminary correlation analyses. Section~\ref{sec:methods} introduces the Bivariate Conditional Poisson (BCP) regression framework, details the log-linear covariate specifications for scoring intensities, and outlines the Bayesian inferential setup using empirical Bayes priors and Hamiltonian Monte Carlo sampling in Stan. Section~\ref{sec:results} reports the empirical findings, evaluating directional specifications ($Home\,\,Goals \rightarrow Away\,\,Goals$ versus $Away \,\,Goals\rightarrow Home\,\, Goals$) via leave-one-out cross-validation, interpreting posterior parameter distributions, and validating model fit through posterior predictive checks. Finally, Section~\ref{sec:conclusion} summarizes the main conclusions, addresses study limitations, and highlights potential avenues for future research.

\section{Data}\label{sec:data}

This section describes the construction of the dataset and presents preliminary empirical patterns that motivate the modelling strategy developed in Section~\ref{sec:methods}.

\subsection{Description and pre-processing}

The dataset comprises match-level observations from three English Premier League (EPL) seasons: 2018--19, 2020--21, and 2023--24. These seasons were selected to represent, respectively, the pre-pandemic period, the period in which most matches were played behind closed doors because of COVID-19 restrictions, and the post-pandemic period. Their inclusion therefore provides useful variation in stadium attendance with which to examine the association between crowd presence and match outcomes.

Match results and statistics were obtained from \url{Football-Data.co.uk}, while match attendance figures were collected from \url{FBref.com}. The two sources were combined using the relevant match identifiers. Each season contains 380 matches, yielding a total sample of 1,140 observations. The observed variables are the numbers of home and away goals (HG and AG), stadium attendance (SA), and the numbers of fouls committed by the home and away teams (HF and AF).

\subsection{Preliminary insights}

Table~\ref{tab:pl_summary} reports selected summary statistics by season. Average attendance fell from 38,181 in the 2018--19 season to only 462 in the COVID-stricken 2020--21 season, reflecting the restrictions imposed during the pandemic, before returning to 38,613 for the 2023--24 season. Over the same pre-pandemic-to-pandemic comparison, mean home goals declined from 1.57 to 1.35, whereas mean away goals increased slightly from 1.25 to 1.34. Although these descriptive differences do not by themselves establish a causal crowd effect, they are consistent with a reduction in home advantage when spectators were largely absent. Scoring was higher for both teams in the 2023--24 season, when mean home and away goals reached 1.80 and 1.48, respectively.

\begin{table}[!htbp]
\centering
\captionsetup{skip=4pt}
\caption{Premier League Summary Statistics by Season}
\begin{tabular}{lrrrr}
\toprule
Season   & Matches & Mean Home Goals & Mean Away Goals & Mean Attendance \\
\midrule
2018--19 & 380     & 1.57            & 1.25            & 38181         \\
2020--21 & 380     & 1.35            & 1.34            &    462          \\
2023--24 & 380     & 1.80            & 1.48            & 38613         \\
\bottomrule
\end{tabular}
\label{tab:pl_summary}
\end{table}

\begin{figure}[H]
    \centering
    \begin{subfigure}[t]{0.49\textwidth}
        \centering
        \includegraphics[width=\linewidth]{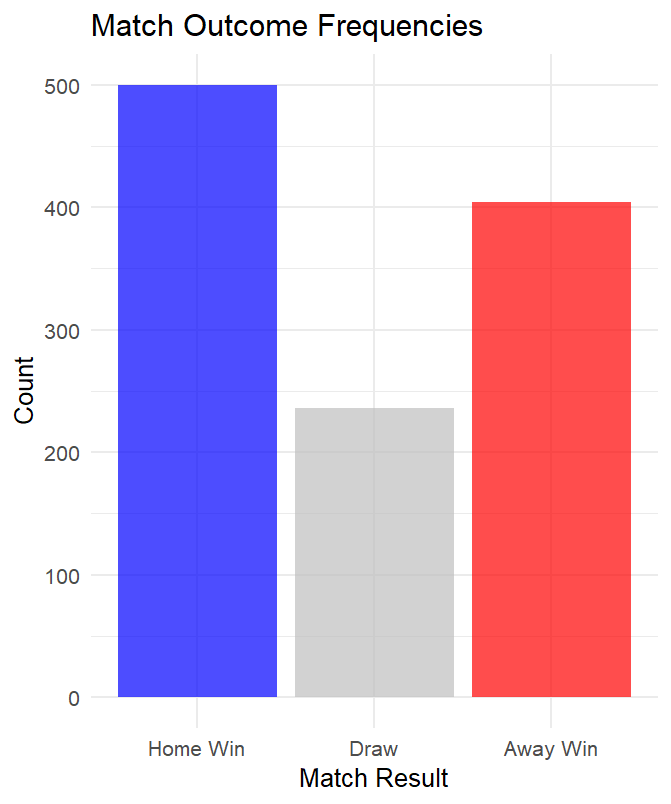}
        \caption{}
        \label{fig:match-outcomes}
    \end{subfigure}
    \hfill
    \begin{subfigure}[t]{0.49\textwidth}
        \centering
        \includegraphics[width=\linewidth]{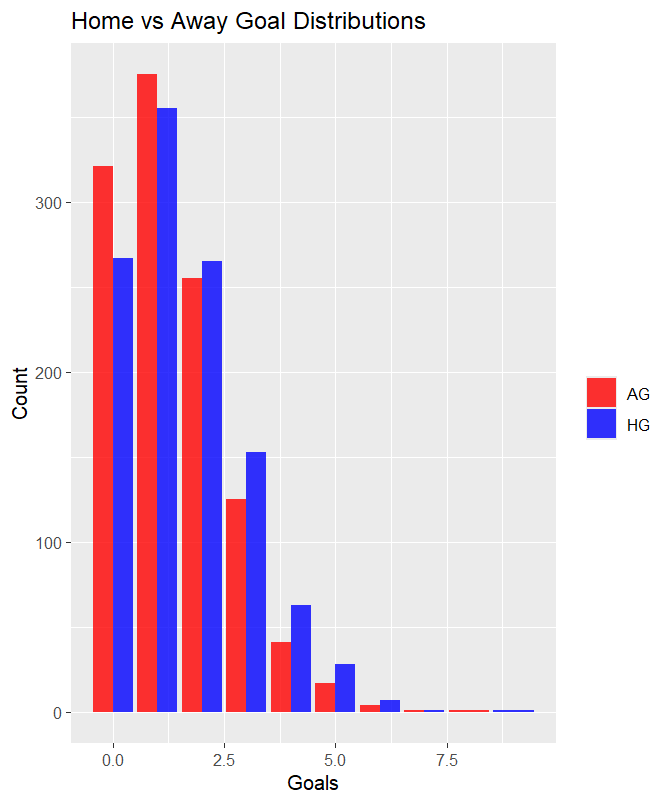}
        \caption{}
        \label{fig:goal-dist}
    \end{subfigure}
    \caption{Distributions of match outcomes (left) and home and away goals (right).}
    \label{fig:match-goal-distributions}
\end{figure}

Across the full sample, home wins are the most frequent match outcome (approximately 500 matches), followed by away wins (approximately 400) and draws (approximately 240); see Figure~\ref{fig:match-outcomes}. This pattern provides descriptive evidence of home advantage and motivates a closer examination of the factors associated with home-team scoring.

Figure~\ref{fig:goal-dist} compares the empirical distributions of home and away goals. Both are right-skewed, with most teams scoring zero, one, or two goals and comparatively few teams recording high scores. This discrete, non-negative structure supports the use of a Poisson-based framework. The home-goal distribution is shifted slightly to the right of the away-goal distribution, in line with the higher average number of goals scored by home teams.

As an initial assessment of the relationship between crowd presence and on-field discipline, Pearson correlations were computed between attendance and the foul measures. The following observations are made. 
\begin{itemize}
    \item Attendance and home fouls were weakly negatively correlated ($r=-0.1097$, $p=0.0002$);
    \item Attendance and away fouls exhibited little evidence of an association ($r=0.0336$, $p=0.2572$); and
    \item Attendance and total fouls were also not significantly correlated at the 5\% level ($r=-0.0499$, $p=0.0924$).
\end{itemize}
These associations are small and should be interpreted descriptively. Nevertheless, they suggest that crowd presence may be related to home-team behaviour while having little direct association with away or total fouls. Together with the interpretation of fouls suffered as a possible proxy for attacking pressure, or for an opponent's inability to halt an attack legally, these patterns motivate the inclusion of foul-related covariates in the regression models.
\section{Methodology} \label{sec:methods}

Let $Y_1$ and $Y_2$ denote the two discrete random variables 
representing the number of goals scored by the two teams in a given football match. A standard starting point in football score modeling is the Poisson distribution. In the simplest case of marginal (independent) Poisson models, it is assumed that $Y_j$ are independent Poisson random variables with means $\mu_j$, i.e., $
Y_j \sim \mathrm{Poisson}(\mu_j), \; j = 1,2$. The mean parameters $\mu_j$ are typically modeled as functions of match-specific covariates or latent team effects. For example, in the well-known Dixon-Coles model \citep{dixon1997}, letting subscripts $H$ and $A$ denote home and away teams respectively, one assumes
\[
Y_H \sim \mathrm{Poisson}(\mu_H), \quad \mu_H = \exp(\gamma + \alpha_H + \beta_A),
\]
\[
Y_A \sim \mathrm{Poisson}(\mu_A), \quad \mu_A = \exp(\alpha_A - \beta_H),
\]
where $\alpha$ and $\beta$ represent attack and defence strength parameters, respectively, and $\gamma$ captures the home advantage effect. Or similarly, in the presence of covariates $x_i$ for match $i$, $Y_{i,j} \sim \mathrm{Poisson}(\mu_{i,j})$, with $\mu_{i,j} = \exp(x_i^\top \beta_j),$
where $\beta_j$ is a vector of regression coefficients.

Note that the above formulation does not model dependence between home and away goals, which is an anticipated feature. Building on this observation, a natural idea would be to consider the bivariate Poisson (BP) distribution. The BP is constructed as follows. Let $Z_0, Z_1, Z_2$ be independent Poisson random variables with parameters $\lambda_0, \lambda_1, \lambda_2$, respectively, and define
$
Y_1 = Z_1 + Z_0, \quad Y_2 = Z_2 + Z_0.
$
The pair $(Y_1, Y_2)$ follows a BP distribution with parameter vector $(\lambda_0, \lambda_1, \lambda_2)$, and marginal distributions that remain Poisson; see \citet{holgate1964,kawamura1984,karlis2003} for some examples. Dependence between $Y_1$ and $Y_2$ is induced through the shared component $Z_0$, yielding $\mathrm{Cov}(Y_1, Y_2) = \lambda_0 \geq 0$. Observe that this particular model formulation can only accommodate non-negative dependence, which is not realistic when modeling competing teams' goal counts at the match level. 

To address this limitation, we develop a Bayesian regression model based on the bivariate conditional Poisson (BCP) distribution of \citet{berkhout2004}, adopting the parameterization of \cite{piancastelli2023} specified in terms of the marginal means and a parameter governing dependence. A (reparameterized) BCP random vector $(Y_1, Y_2)$, denoted $(Y_1, Y_2) \sim \mathrm{BCP}(\lambda_1, \lambda_2, \phi)$, admits the representation
\begin{eqnarray}\label{storep}
Y_1 \sim \mathrm{Poisson}(\lambda_1), \quad 
Y_2 \mid Y_1 = y_1 \sim \mathrm{Poisson}\big(\mu_2 e^{\phi y_1}\big),
\end{eqnarray}
where $\mu_2 = \lambda_2 \exp\{-\lambda_1(e^{\phi}-1)\}, \lambda_1, \lambda_2 >0$ and $\phi \in \mathbb{R}$. We denote $(Y_1,Y_2)\sim\mbox{BCP}(\lambda_1,\lambda_2,\phi)$. Notably, the BCP model introduces a directional structure because the distribution of $Y_2$ is conditional on $Y_1$. In our context, the assignment of home and away goals to $Y_1$ or $Y_2$ reflects a modeling choice regarding which component is assumed to drive the other. Both specifications, home-to-away and away-to-home, will be considered and compared in Section \ref{sec:results}.

The joint probability function of a BCP random pair $(Y_1, Y_2)$, denoted by $f(y_1,y_2|\lambda_1, \lambda_2, \phi) \equiv \mathbb{P}(Y_1=y_1, Y_2=y_2)$, is given by
\[
f(y_1,y_2|\lambda_1, \lambda_2, \phi) = \frac{\lambda_1^{y_1} \lambda_2^{y_2}}{y_1!y_2!}
\exp\left\{
-\lambda_1 \big(1 + y_2(e^{\phi}-1)\big)
- \lambda_2 \exp\{-\lambda_1(e^{\phi}-1) + \phi y_1\}
+ \phi y_1 y_2
\right\},
\]
for $(y_1,y_2)\in\mathbb\{0,1,\ldots\}^2$. Naturally, as $Y_1 \sim \mbox{Poisson}(\lambda_1)$, its marginal mean and variance are $\mathbb{E}(Y_1) = \mathrm{Var}(Y_1) = \lambda_1$. As for $Y_2$, its (unconditional) distribution is a mixed Poisson with $\mathbb{E}(Y_2) = \lambda_2$ and $\mathrm{Var}(Y_2) = \lambda_2 + \lambda_1 \lambda_2 (e^{\phi}-1)^2.$ The BCP covariance and correlation are respectively given by \citep{piancastelli2023}
\[
\mathrm{Cov}(Y_1, Y_2) = \lambda_1 \lambda_2 \bigl(e^{\phi} - 1\bigr),
\]
and
\[
\mathrm{Corr}(Y_1, Y_2) = \bigl(e^{\phi} - 1\bigr)\sqrt{\dfrac{\lambda_1 \lambda_2 }{1+\lambda_2(e^{\lambda_1(e^\phi-1)^2}-1)}},
\]
which implies that the direction of dependence is determined by the sign of $\phi$. Illustrations pertaining to the role of the $\phi$ parameter in $\mathrm{Corr}(Y_1, Y_2)$ can be found in \cite{piancastelli_ingarch} (Section 2). In particular, we note that the correlation structure as a function of $\phi$ is nonlinear, with values of $\phi$ closer to zero corresponding to stronger dependence even though $\phi=0$ is the independence case. Further, the model is able to account for an extended range of positive or negative correlations. More properties of the BCP distribution, such as joint moments, can be found in \cite{berkhout2004}.

We now propose the BCP to model football scoring outcomes. Let $\boldsymbol{y}_i = (y_{i,1}, y_{i,2})^\top$ represent the goals scored by each team in match $i$, and assume that ${y}_i \sim \text{BCP}(\lambda_{1,i}, \lambda_{2,i},\phi)$. The specific assignment of home-to-away ($Y_1$: Home, $Y_2:$ Away) and away-to-home ($Y_1$: Away, $Y_2:$ Home) are explored in Section \ref{sec:results}. 

The collection of results across $n$ matches will be denoted by $\boldsymbol{y} = (\boldsymbol{y}_1, \dots, \boldsymbol{y}_n)^\top$. 
To account for the match-specific covariates introduced in Section \ref{sec:data},  log-linear specifications are introduced for $\lambda_{1,i}, \lambda_{i,2}$. Specifically, 
\begin{equation}\nonumber
    \log\lambda_{1,i} = \beta_{0,1} + \beta_{1, Att} \log(Att_i +1) + \beta_{1, F} FS_{i,1},
\end{equation}
\begin{equation}\nonumber
    \log\lambda_{2,i} = \beta_{0,2} + \beta_{2, Att} \log(Att_i +1) + \beta_{2, F} FS_{i,2},
\end{equation}
where $FS_{i,j}$ refers to the fouls suffered by team $j$ in match $i$, and  $Att_i$ the corresponding stadium attendance covariate, with the $\beta$'s denoting the associate regression coefficients. The attendance covariate is log-transformed to manage numerical scaling, and the unit shift ($\log(\text{Att}_i + 1)$) ensures the transformation is well-defined during periods of zero attendance, i.e. closed doors matches that occurred during the 2020-21 season. In what follows, $X$ will denote the design matrix containing these three ($FS_{1}, FS_2, \log(Att +1)$) covariates.

Assuming that the matches are independent, the likelihood function for the collection of results $\boldsymbol{y}$ is the product 
\begin{align} \label{eq:cpois_likelihood}
    f(\boldsymbol{y}| \boldsymbol{\beta}, \phi, X) = \prod_{i=1}^n \Biggl( \frac{\lambda_{1,i}^{y_{i,1}} \lambda_{2,i}^{y_{i,2}}}{y_{i,1}! y_{i,2}!} \exp \Bigl\{ -\lambda_{1,i} \big(1 + y_{i,2}(e^{\phi}-1)\big)  
     - \lambda_{2,i} \exp \{-\lambda_{1,i}(e^{\phi}-1) + \phi y_{i,1} \} + \phi y_{i,1} y_{i,2} \Bigr\} \Biggr),
\end{align}
where $\boldsymbol{\beta} = (\beta_{0,1}, \beta_{1, Att}, \beta_{1, F}, \beta_{0,2}, \beta_{2, Att}, \beta_{2, F})^\top$ denotes the vector coefficients corresponding to the covariates, and $\theta = (\boldsymbol{\beta}^\top, \phi)^\top$ denotes the entire parameter vector of interest. 

Before presenting the Bayesian inferential framework for BCP regression in the next subsection, we highlight key differences between our methodology and the approach proposed by \cite{petetal2025}. Their work also introduces a bivariate conditional Poisson regression inspired by \citet{berkhout2004}, but assumes a baseline joint distribution with the following reparameterized stochastic representation:
\begin{eqnarray}\label{storep2}
Y_1 \sim \mathrm{Poisson}(\lambda_1), \quad 
Y_2 \mid Y_1 = y_1 \sim \mathrm{Poisson}\big(\lambda_2 e^{\phi \operatorname{logit}(F(y_1))}\big),
\end{eqnarray}
where $\lambda_1,\lambda_2>0$, $\phi\in\mathbb R$, $\operatorname{logit}(x)=\log\left(\dfrac{x}{1-x}\right)$ for $x\in(0,1)$, and $F(\cdot)$ being the cumulative distribution function of $Y_1$. The primary distinction between stochastic representations \eqref{storep} and \eqref{storep2} lies in their respective dependence components, $e^{\phi y_1}$ and $e^{\phi \operatorname{logit}(F(y_1))}$. Under the former, mathematical tractability is preserved: key theoretical properties, including marginal means, variances, and the attainable correlation range, are available in closed form and well documented \citep{berkhout2004, piancastelli2023}. Under the latter, \cite{petetal2025} additionally consider a mixture of bivariate conditional Poisson distributions by swapping the roles of $Y_1$ and $Y_2$. However, as those authors acknowledge, the resulting marginal distributions lack finite moments, an unappealing theoretical property when modeling goal counts, which naturally possess finite moments. Beyond these analytical differences, the two approaches also diverge in implementation: while \cite{petetal2025} adopt a frequentist estimation scheme with a distinct set of covariates, our work employs a Bayesian inferential framework.


\subsection{Bayesian Inference }\label{sec:bayes}

We here develop a Bayesian inference for the BCP regression model. Key advantages of adopting a Bayesian approach include directly interpretable uncertainty quantification of model parameters and the capacity to formally incorporate expert knowledge through the prior.

Given the BCP likelihood defined in \eqref{eq:cpois_likelihood}, the  target posterior distribution is given by
\begin{equation}\label{eq:posterior}
    \pi(\boldsymbol{\beta}, \phi | \boldsymbol{y}, X) \propto f(\boldsymbol{y}| \boldsymbol{\beta}, \phi, X) \pi(\boldsymbol{\beta}) p(\phi),
\end{equation}
where $\pi(\boldsymbol{\beta})$ and $p(\phi)$ denote the prior distribution specifications for $\boldsymbol{\beta}$ and $\phi$, respectively.

A number of approaches exist for choosing the prior distributions, such as (i) \textit{subjective priors} - which encode expert belief about the model parameters, (ii) \textit{objective (or noninformative) priors} - designed to have minimal influence on the posterior, and (iii) \textit{empirical Bayes} - wherein parameters of prior distributions are set to certain empirical estimates obtained from the data. In this work, we adopt the empirical Bayes approach, more specifically, letting the frequentist maximum likelihood estimates (MLEs) inform the $\pi(\cdot)$ distributional setup as follows; see \citet{carlin2000}. 

We start from the canonical prior choice for the real-valued parameters, a zero-mean Gaussian distribution. Let $\theta \in \Theta$ denote the model parameter vector, $\hat{\theta}$ its corresponding MLE, and $\widehat{\text{se}}(\hat{\theta})$ the associated frequentist standard error. We specify the prior for $\theta$ as $\theta \sim \mathcal{N}(0, \widehat{\text{se}}(\hat{\theta})^2)$ in an empirical Bayes fashion. Centering the distribution at zero allows the prior to remain weakly informative in the sense of \citet{gelman2013}, allowing the data to dominate posterior inference. Meanwhile, scaling the prior standard deviation using empirical standard errors yields a hybrid prior that is empirically guided. Finally, we apply this empirical approach under together with the assumption of prior independence for $\boldsymbol{\theta} = (\boldsymbol{\beta}, \phi)$, so $\pi(\boldsymbol{\theta}) = \pi(\phi) \prod_{j=1}^{2} \prod_{k \in \{0, Att, F\}} \pi(\beta_{j,k})$.

As is common in most non-trivial Bayesian models, the form of the posterior in  \eqref{eq:posterior} is not known analytically, nor is it available in closed-form. Specifically, its normalizing constant, often referred to as the \textit{marginal likelihood} or \textit{evidence}, is the high-dimensional integral  $ p(\boldsymbol{y} | X) = \int_{\Theta} f(\boldsymbol{y}| \boldsymbol{\theta}, X) \pi(\boldsymbol{\theta}) d\boldsymbol{\theta}$, 
rendering the posterior density intractable.

To circumvent this, we adopt computational simulation methods as done typically. Markov Chain Monte Carlo (MCMC) algorithms are ubiquitous in Bayesian inference; indeed, their advancement has been a crucial driver for the wider adoption of Bayesian methods. Rather than seeking a closed-form analytical solution for the posterior, MCMC techniques generate a sequence of samples from the parameter space. This is achieved by constructing a Markov chain designed to have the target posterior $\pi(\boldsymbol{\theta} | \boldsymbol{y}, X)$ as its stationary distribution. Crucially, because MCMC algorithms typically rely on ratios of posterior densities, such as the Metropolis-Hastings ratio, the normalizing constant $p(\boldsymbol{y} | X)$ is avoided. Once the chain has reached its stationary region, the resulting samples can be treated as samples from the posterior and used to approximate summaries of interest.

We perform MCMC sampling using Stan \citep{stan}, a software which implements a computationally efficient MCMC algorithm called Hamiltonian Monte Carlo (HMC). A feature of HMC is that it uses gradient information to explore high-probability regions of the posterior effectively. Details on this sampler are provided in \citet{hoffman2014}. Our code for implementing the BCP model in Stan is provided by the authors upon request. 

\subsection{Posterior predictive assessment}

Another advantage of Bayesian inference is the ability to generate samples from the posterior predictive distribution. This represents the distribution of new simulated and replicated data ($y_{rep}$) that could be expected given the observed data ($y$). It is formally defined by integrating the likelihood over the entire posterior distribution of the parameters. $$p(y_{rep} | y) = \int p(y_{rep} | \theta) p(\theta | y) d\theta$$ This is accomplished by first drawing a parameter sample, $\theta^{(s)}$, from the MCMC generated posterior, and then drawing a new dataset, $y_{rep}^{(s)}$, from the likelihood function conditional on that parameter sample. This process is automated in Stan. This results in a collection of replicated datasets that can be used for model checking. It also allows for generating predictions that allow interpretation of parameter uncertainty. \citet{gelman1996} outline in depth the theory on Posterior Predictive assessment.

Posterior Predictive Checks (PPCs) are used to assess the model's ability to match the observed data. This is achieved by comparing the distribution of the observed data with the distributions of replicated datasets. Graphical comparisons are assessed. Density plots of replicated data on the observed data and comparing summary statistics between the observed and replicated datasets are analysed. A model that fits well will generate replicated data that looks similar to the observed data.

\subsection{Model selection}

The relative predictive performance of Bayesian models is compared using information criteria that estimate out-of-sample predictive accuracy while penalizing for model complexity. The Widely Applicable Information Criterion (WAIC) and the Leave-One-Out Cross-Validation (LOO-CV), implemented via the `loo` package, are calculated. Models with lower WAIC or LOO-CV values are considered to have better predictive performance. \citet{vehtari2017} outline the relevant theory. The Expected Log-Predictive Density (ELPD) is also calculated for model comparison of the Bayesian methods to quantify and compare the out-of-sample predictive power of the conditional vs the independent model. The larger difference favours the model with the higher ELPD score.

Beyond purely statistical measures, the final model selection is guided by the interpretability and theoretical coherence of the parameter estimates. The primary focus is on how competing models capture the influence of the covariates and the dependence parameter $\phi$. The main goal is to select a model that provides a statistically valid and contextually meaningful explanation of the data.


\section{Results}\label{sec:results}

In this section, we analyze the English Premier League (EPL) data introduced in Section \ref{sec:data} with the Bayesian conditional bivariate Poisson (BCP) regression specified in Section \ref{sec:methods}. We study the two possible direction attributions for the BCP: one in which the away team goals influence the home team  goals (A$\to$H), and the other in which the home team goals influence the away team goals (H$\to$A). Bayesian inference is conducted in R using the \texttt{rstan} interface to the Stan probabilistic programming platform. Sampling was performed across four chains, each running 2,000 iterations, with the first 1,000 iterations discarded as warm-up. Empirical Bayes priors specified in Section \ref{sec:bayes} are employed, although results are robust to the prior choice as investigated by the authors but not reported here to save space.

The following three models are fitted: (i) BCP A$\to$H, (ii) BCP H$\to$A, and (iii) independent Poisson, and first compared via their leave-one-out expected log predictive density (ELPD-LOO) performance. ELPD-LOO is a fully Bayesian selection criterion that quantifies how well a model predicts unseen data. Its computation involves evaluating the expected log posterior of each sample point $i$, i.e., match $i$, under a model that is fit without data from match $i$. Formally, 
\begin{equation*}
    \text{ELPD}_i = \mathbb{E}_{\theta \sim p(\theta \mid y_{-i})} \big[ \log p(y_i \mid \theta) \big],
\end{equation*}
where $p(\theta \mid y_{-i})$ is the posterior distribution of the parameter vector $\theta$ obtained by fitting the model without the observations from match $i$. The total ELPD-LOO is then $\text{ELPD}_\text{LOO} = \sum_{i=1}^{n} \text{ELPD}_i$. In practice, exact leave-one-out computation is approximated using Pareto-smoothed importance sampling (PSIS-LOO) to avoid $n$ model refits. Higher ELPD-LOO values are indicative of better out-of-sample predictive performance. The ELPD-LOO estimates were BCP A$\to$H: $-3552.0$, BCP H$\to$A: $-3552.1$ and Poisson: $-3565.0$. These results point to similar performance shown by the two BCP models, with some indication of improvement when compared to the independent Poisson model. 

Next, we study the directionality choice with ELPD-LOO metric by considering the conditional log-probability of the causal component, i.e., $\text{ELPD}_{2} = \sum_{i=1}^n
\text{ELPD}_{i,2}$ where $\text{ELPD}_{i,2} = \mathbb{E}_{\theta \sim p(\theta \mid {y}_{2, -i}, {y}_{1})}$  $\big[ \log p(y_{i,2} \mid y_{i,1}, \theta) \big]$. The H$\to$A attribution yields an $\text{ELPD}_{2}$ estimate of $-1721.8$ and the A$\to$H direction yields $-1817.2$. Hence, while both models achieve similar joint predictive performance, conditional predictive evaluation shows that the H$\to$A model provides better predictive accuracy, supporting this specification.

\begin{figure}[ht]
\centering
\begin{minipage}[t]{0.4\textwidth} 
\vspace{0pt} 
\centering
\captionsetup{type=table}
\renewcommand{\arraystretch}{1.3} 
\begin{tabular}{lcc}
\hline
 & Mean (SD) & 95\% CI \\
\hline
$\beta_1^0$   & 0.146 (0.084) & [$-$0.022, 0.311] \\
$\beta_{1,\text{Att}}$ & 0.024 (0.005) & [0.013, 0.034] \\
$\beta_{1,\text{AF}}$  & 0.012 (0.007) & [$-$0.001, 0.025] \\
$\beta_2^0$   & 0.232 (0.096) & [0.041, 0.419] \\
$\beta_{2,\text{Att}}$ & 0.000 (0.006) & [$-$0.011, 0.011] \\
$\beta_{2,\text{HF}}$  & 0.007 (0.007) & [$-$0.008, 0.022] \\
$\phi$        & $-$0.107 (0.020) & [$-$0.147, $-$0.066] \\
\hline
\end{tabular}
\caption{Posterior summaries of the BCP H$\to$A model parameters. Means, standard deviations (SD), and 95\% credible intervals (CI) are reported for all parameters.} \label{tab:posterior}
\end{minipage}%
\hspace{0.08\textwidth}
\begin{minipage}[t]{0.42\textwidth} 
\vspace{0pt} 
\centering
\includegraphics[width=\textwidth]{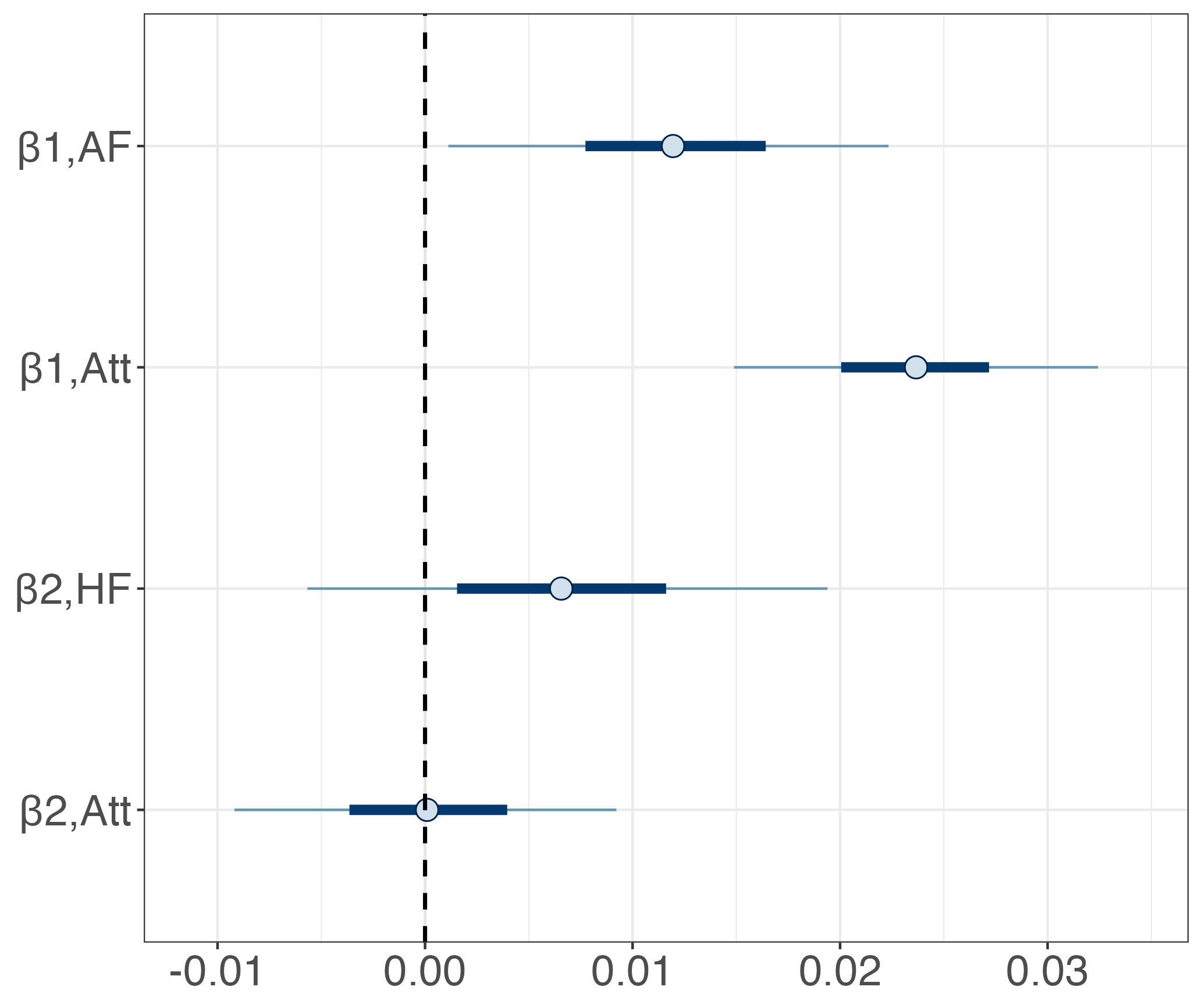}
\caption{Posterior intervals for covariate effects. Thick bars represent 50\% intervals, thin bars 90\% intervals, and points indicate medians.} \label{fig:betas}
\end{minipage}
\end{figure}

The posterior estimates from the BCP H$\to$A model are summarized in Table~\ref{tab:posterior}. The validity of posterior inference was assessed using trace plots, the Gelman--Rubin statistic ($\hat{R}$; \cite{gelman2013}), and the effective sample size ($n_{\text{eff}}$). The trace plots show no apparent trends or drifts, indicating that all four chains adequately explored the target distribution; these plots are omitted here to save space. In addition, $\hat{R}$ values are close to 1.0 for all parameters, and effective sample sizes are sufficiently large.

As described in Section~\ref{sec:methods}, the parameter $\phi$ captures the dependency between home and away team goals, with its sign indicating the direction of correlation and $\phi = 0$ representing independence. The posterior for $\phi$ is centered at $-0.107$, with a 95\% credible interval that excludes zero, indicating a statistically significant negative relationship. This suggests a suppressive effect in which goals scored by the home team are associated with a reduction in the expected number of away team goals, consistent with theoretical expectations.

A visualisation of the posterior intervals corresponding to the covariate effects is provided in Figure~\ref{fig:betas}. Thick bars represent 50\% credible intervals, thin bars 90\% credible intervals, and points indicate the posterior medians. We observe that the effect of (log) match attendance on home team goals, represented by $\beta_{1,\mathrm{Att}}$, is positive and significant, whereas the corresponding effect on away team goals, $\beta_{2,\mathrm{Att}}$, is centered at zero. This result empirically confirms the well-known influence of crowd support on the performance of the home team. The effect of fouls is also not symmetric. While the results on the parameter $\beta_{1,\mathrm{AF}}$ indicate that away team fouls positively impact home team goals, the reverse is not decisive. Specifically, the 95\% credible bands for $\beta_{2,\mathrm{HF}}$ include zero. These findings highlight the asymmetric impact of contextual and behavioral factors on team performance during matches. Such asymmetry strongly supports the home advantage phenomenon, whereby the home team benefits from both crowd support and opposition disruptions.

The fit of the BCP H$\to$A model to the data is further assessed using a posterior predictive exercise. Posterior predictive checks (PPCs) are a standard Bayesian tool for evaluating model adequacy, based on comparing the observed data to data replicated from the fitted model. Specifically, parameter values are sampled from the posterior distribution, and replicated datasets are generated from the likelihood conditional on these sampled parameters. These replicated datasets are compared to the observed data, typically through a set of informative summary statistics. The choice of summary statistics is context-dependent and aims to assess how well the model captures key features of the data. We consider the marginal mean and variance of home and away team goals, the sample correlation between goals, and the proportion of matches lost by the home team.

\begin{figure}
    \centering
    \includegraphics[width=0.85\linewidth]{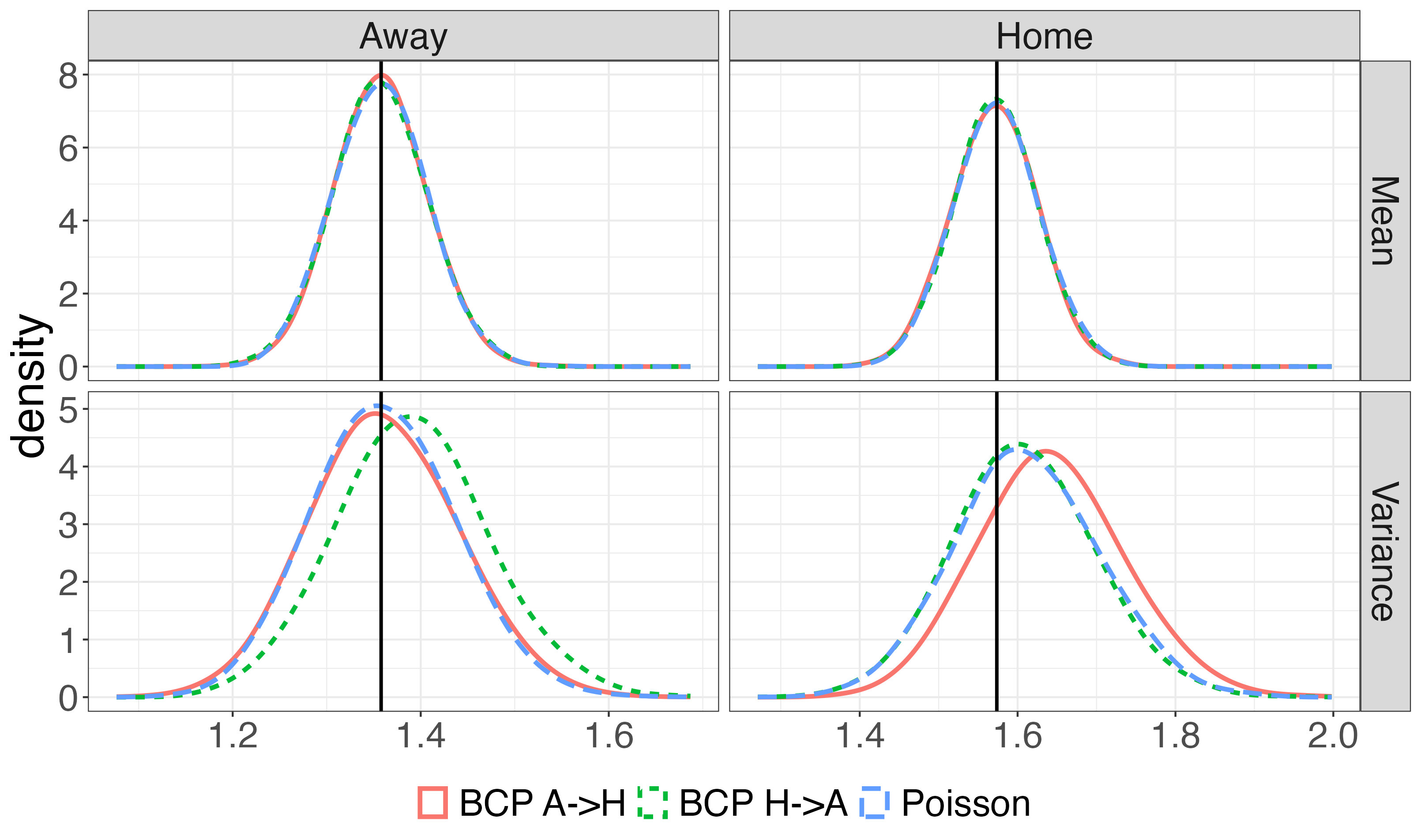}
\caption{Posterior predictive densities for the marginal means and variances of goals. Teams are represented as columns and statistics as rows. Vertical solid lines indicate the observed value.}\label{fig:ppc1}
\end{figure}

\begin{figure}
    \centering
    \includegraphics[width=0.85\linewidth]{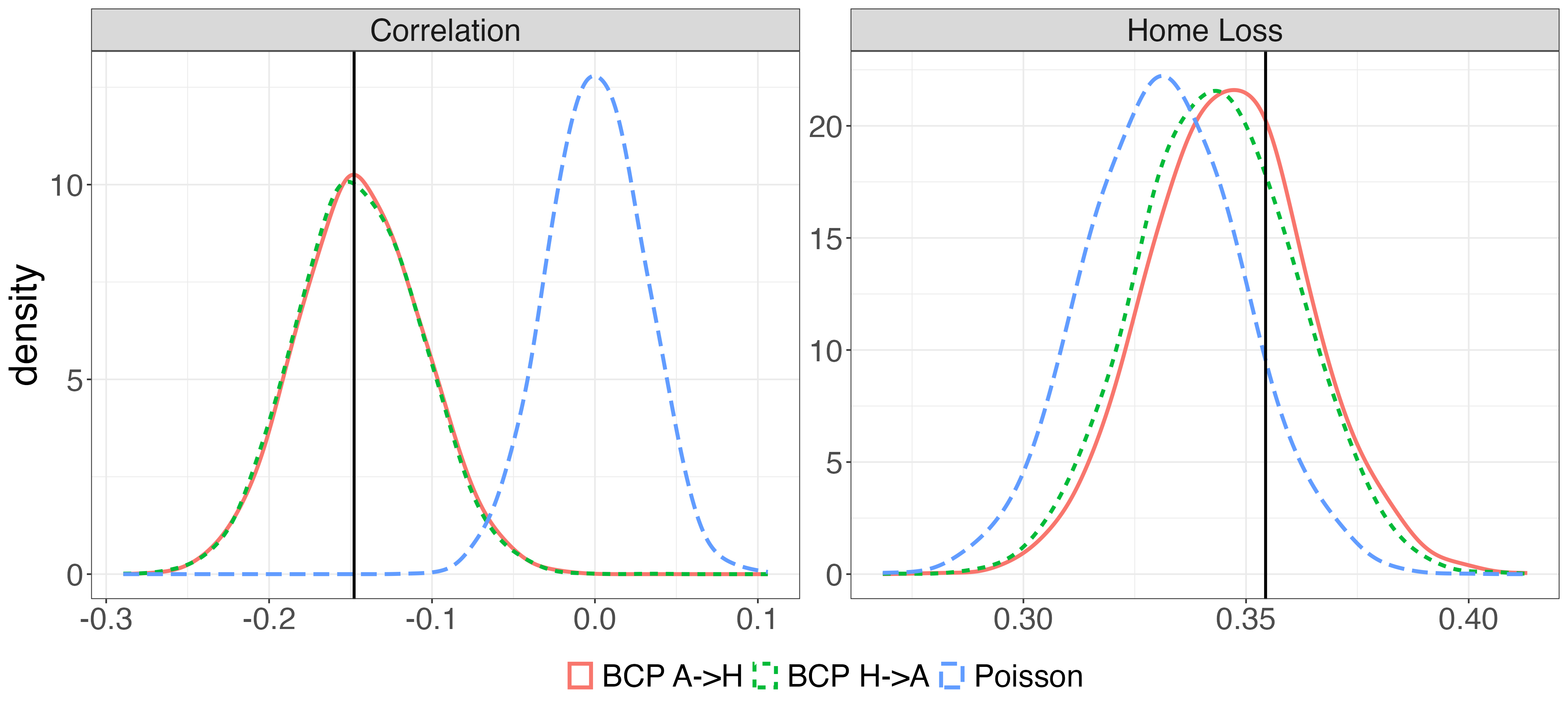}
  \caption{Posterior predictive densities for the correlation between home and away goals (left) and the proportion of matches lost by the home team (right). } \label{fig:ppc2}
\end{figure}

Posterior predictive checks are repeated for all three competing models, BCP H$\to$A, BCP A$\to$H, and the independent Poisson model, as this comparison is expected to highlight important distinctions. Figures \ref{fig:ppc1} and \ref{fig:ppc2} illustrate the posterior predictive results with densities constructed from 4,000 replications. Models are indicated by different colors and line types, while the vertical solid line represents the observed value of each summary statistic in the data. Figure \ref{fig:ppc1} shows the marginal means and variances, with teams as columns and statistics as rows. The observed values lie within high posterior predictive density regions for all models, indicating that all adequately capture the marginal mean and variance structures. However, this changes when assessing the goals correlation and the proportion of home losses, shown in Figure \ref{fig:ppc2}. The observed dependency between goals is not captured by the independent Poisson model, consistent with our previous observation on the significance of $\phi$. Incorporating this dependency leads to a more realistic representation of match outcomes, with the BCP models producing home loss proportions closer to those observed in the data.

\subsection{Discussion}
\label{sec:discussion}

The model comparisons provide evidence that allowing for negative dependence between home and away team goals improves the representation of match-level dynamics from the EPL. The two BCP specifications achieved nearly identical joint ELPD-LOO values ($-3552.0$ for A$\to$H and $-3552.1$ for H$\to$A), with both models seen doing better than the independent Poisson model. Thus, this joint ELPD-LOO criterion supports the inclusion of a dependence structure but does not distinguish meaningfully between the two directional factorizations. By contrast, the conditional ELPD comparison favored H$\to$A ($-1721.8$ versus $-1817.2$), motivating our focus on that specification. This direction should be interpreted as a predictive factorization of the joint distribution, however, rather than as evidence that home team goals causally determine away team goals.

The posterior estimate of the dependence parameter provides clear evidence of a negative association between the two scoring processes. Under the H$\to$A specification, the posterior mean of $\phi$ is $-0.107$, and its 95\% credible interval, $[-0.147,-0.066]$, excludes zero. Because the conditional mean of away team goals contains the factor $\exp(\phi y_{1})$, a one-goal increase in the home score multiplies the conditional expected number of away goals by $\exp(-0.107)\approx 0.899$, holding the modeled intensities and covariates fixed. This corresponds to an estimated reduction of approximately 10\% in the conditional mean, not a 10\% reduction in the probability of scoring. The negative dependence may reflect tactical and behavioral adjustments within a match, such as a team protecting a lead or its opponent changing its level of attacking risk.

The results highlight an important advantage of the conditional construction of \citet{berkhout2004}. Conventional shared-component bivariate Poisson models restrict the covariance to be non-negative, whereas the BCP model accommodates both signs of association. The posterior predictive checks reinforce the practical importance of this flexibility. Specifically, although all three models reproduce the marginal means and variances reasonably well, the independent Poisson model fails to reproduce the observed goal correlation and yields a less accurate distribution for the proportion of home losses. The BCP models therefore provide a more realistic joint description of the match outcomes considered here.

The covariate results also indicate a pronounced asymmetry between home and away scoring. The coefficient of log attendance is positive for home team goals, with a posterior mean of $0.024$ and a 95\% credible interval of $[0.013,0.034]$, whereas the corresponding interval for away team goals, $[-0.011,0.011]$, is centered at zero. Because attendance enters as the term $\log(\mathrm{Attendance}+1)$ in a log-linear mean model, the home goal coefficient can be interpreted approximately as an elasticity at positive attendance levels. More precisely, a 1\% increase in attendance is associated with about a 0.024\% increase in the expected number of home goals, with the other modeled quantities held fixed. The home-specific pattern is consistent with previous evidence that the presence of crowds primarily affects home-team scoring \citep{piancastelli2023}. Nevertheless, attendance, pandemic conditions, and season are closely intertwined in these data, so the coefficient should be interpreted as an adjusted association rather than a causal crowd effect.

Evidence for the foul-related covariates is also asymmetric. The posterior mean for the association between away team fouls and home team goals is positive ($0.012$), but its 95\% credible interval, $[-0.001,0.025]$, includes zero. Likewise, the interval for the association between home team fouls and away team goals, $[-0.008,0.022]$, includes zero. The data therefore suggest, at most, a weak positive association between away fouls and home scoring; they do not establish a decisive foul effect at the 95\% credible level. More detailed measures of match events and team strength may be needed to separate the roles of attacking pressure, defensive disruption, and refereeing behavior.


\section{Conclusion}
\label{sec:conclusion}

This paper developed a Bayesian bivariate conditional Poisson model to analyse 1,140 English Premier League matches from the 2018--19, 2020--21, and 2023--24 seasons. Extending the established Poisson-based literature on football scores \citep{maher1982,dixon1997,karlis2003}, the BCP formulation allows the association between home and away goals to be either positive or negative while retaining separate regression structures for the two scoring intensities. Bayesian estimation in Stan provides direct uncertainty quantification for both the dependence parameter and the covariate effects.

The results support three main conclusions. First, the BCP models provide better joint predictive performance than the independent Poisson model and more accurately reproduce the observed correlation between home and away goals. Second, the preferred H$\to$A specification yields a negative dependence estimate, with the credible interval for $\phi$ lying entirely below zero. Third, attendance is positively associated with home-team scoring but shows no clear association with away-team scoring, a pattern consistent with earlier research on crowd presence and home advantage \citep{piancastelli2023}. The foul-related effects are comparatively uncertain because both 95\% credible intervals include zero.

These findings should be interpreted in light of the study design. The three selected seasons provide substantial variation in attendance, but they also differ in pandemic conditions, team composition, and the broader scoring environment. Moreover, the current specification does not explicitly account for team-specific attacking and defensive strengths. The estimated attendance effect therefore describes an association conditional on the included covariates and should not, by itself, be interpreted as a causal effect of crowd size.

Several extensions would strengthen the analysis. Hierarchical team-level attack and defense effects, together with season-specific or time-varying parameters, could account for changes in team quality and scoring conditions; related seasonal structures are considered by \citet{egidi2018}. Additional match-level information, including expected goals, red cards, rest periods, tactical formations, weather, and referee characteristics, could help explain the mechanisms underlying the estimated dependence. Applying the model to more seasons and to leagues in other countries would also show whether the negative association and attendance asymmetry generalize across competitions and playing styles. Finally, designs that compare matches within narrower time windows or exploit more localized changes in spectator restrictions would permit a more credible assessment of causal crowd effects.

\paragraph{Software and data availability.}
The Stan implementation of the BCP model can be obtained from the authors upon request. The original match statistics and attendance records were obtained from the publicly accessible sources identified in Section~\ref{sec:data}; the data-construction steps needed to reproduce the analytical dataset are described there.

\paragraph{Conflict of interest.} None to be declared.

\end{document}